\documentstyle[12pt]{article}
\begin{document}
\hoffset=-1truecm
\voffset=-2truecm
\baselineskip=18pt plus 2pt minus 2pt
\hbadness=500
\tolerance=5000
\title{\bf Reply to ``Comment on `Two Fermi points in the exact
solution of the Kondo problem''}
\author{A. A. Zvyagin \\
Max-Planck-Institut f\"ur Chemische Physik fester Stoffe, \\
N\"othnitzer Str. 40, D-01187 Dresden, Germany, and \\
B.I. Verkin Institute for Low Temperature Physics and Engineering, \\
Ukrainian National Academy of Sciences, 47 Lenin Avenue, \\
Kharkov, 61164, Ukraine \\}
\date{December 14, 2001}
\maketitle
\begin{abstract}
In his Comment N.~Andrei questions the use of symmetric limits of
integration of the Bethe ansatz integral equations in my recent study
devoted to some features of the Bethe ansatz solution of the Kondo
problem. In this Reply I show that the statement of the Comment about the
asymmetry of integration limits contradicts the distribution of the
quantum numbers in discrete Bethe ansatz equations. I also argue that
the asymmetry of the excitation energy, supported in the Comment,
contradicts the initial chiral symmetry of conduction electrons in 
the physical Kondo problem. Hence the distribution of spin rapidities
has to be symmetric for any magnetization. 
\end{abstract}

\newpage

The Comment by N.~Andrei \cite{And} is devoted to the discussion of
the very important point of the Kondo problem. Recently I pointed out
that two energy scales can appear in the Bethe ansatz solution of the
Kondo problem \cite{Zv1}. The onset of the second scale was the
consequence of the symmetric with respect to zero distribution of 
spin rapidities, which parametrize the eigenvalues and eigenstates of
the Hamiltonian of the Kondo problem in the Bethe ansatz approach. The
main question raised by the Comment is whether that distribution of
spin rapidities is symmetric or asymmetric. In other words, the
Comment questions the use of symmetric limits of integration of the
Bethe ansatz integral equations in Ref.~\cite{Zv1}. I note that
for the presence of two energy scales in the Bethe ansatz solution it
is enough to consider both of limits being not necessary symmetric,
but non-infinite, though. One infinite and one finite limits of
integration produce ill-defined features of the ground states of the
impurity models in the Bethe ansatz solutions and it was necessary to 
introduce special artificial cut-off procedures to avoid these
problems \cite{obz}.  

There are two arguments that support the symmetric limits of
integration of my approach: 

1. In the discrete Bethe ansatz equations for spin rapidities
[Eqs.~(1) of the Comment or Eqs.~(3) of Ref.~\cite{Zv1}] the
quantum numbers $J_{\alpha}$, which parametrize the eigenstates and
eigenvalues are distributed {\em symmetrically} with respect to zero,
between $\pm (N-M-1)/2$ \cite{And,Zv1}. Here $N$ is the number of
electrons and $M$ is the number of electrons with down spin (including
the impurity). Hence, this symmeric distribution is valid for {\em
any} $M$, i.e., for any magnetic moment of the system [the
$z$-projection of the total spin of the system is equal to $S^z =
(N/2) -M$]. The integral Bethe ansatz equations (Eq.~(4) of the
Comment and Eq.~(4) of Ref.~\cite{Zv1}) are the direct
consequences of discrete Bethe ansatz equations (Eqs.~(1) of the
Comment or Eqs.~(3) of Ref.~\cite{Zv1}), according to the 
well-known procedure \cite{Hul}. Hence the distribution of rapidities
in the thermodynamic continuum limit must be also symmetric for any 
magnetization of the system, otherwise the distributions of quantum
numbers in the discrete Bethe ansatz equations and in their continuum
limit would contradict each other.  
 
2. The energy of any excitation, mentioned in the Comment, is,
naturally, the consequence of the Bethe ansatz equations for quantum
numbers (or for the distribution of rapidities). One cannot obtain
that energy independently of the Bethe ansatz equations. N.~Andrei
points out that this energy is the asymmetric function of spin
rapidities in the known Bethe ansatz solution of the Kondo problem. I
argue that this is the artifact of the approach. Physically, one can
consider both signs for the kinetic term in the Hamiltonian of the
problem (the first equation of the Comment and Eq.~(1) of 
Ref.~\cite{Zv1}) because one can study the Kondo interaction of 
the magnetic impurity with either right- or left-moving conduction
electrons from the same grounds. The physical answer must not depend
on the choise of the chirality of conduction electrons,
naturally. However, if one uses the definition of the energy supported
in the Comment, the energy changes its sign when changing the
chirality of conduction electrons, namely due to that
asymmetry. Clearly, one has to use the symmetrized energy, too, to
avoid this inconsistency. That implies that the energy should be defined as 
$E = -\sum_{j=1}^N |k_j| + {\rm const}$ \cite{er}. 
This removes the inconsistency between the symmetric distribution of
quantum numbers in the discrete Bethe ansatz equations and the
asymmetric distribution in its continous limit [Eq.~(4) of the
Comment], mentioned above. It turn out that, emphasized in
the Comment, distributions of rapidities and energies are symmetric 
functions with respect to zero in any known Bethe ansatz solutions 
(lattice ones or quantum field theories) {\em except of impurity problems}. 

Summarizing, the statement of the Comment about the asymmetry of 
integration limits contradicts the distribution of the quantum numbers
in discrete Bethe ansatz equations. On the other hand, the asymmetry
of the excitation energy supported in the Comment, contradicts the
initial chiral symmetry of conduction electrons in the physical Kondo 
problem. Therefore the distribution of spin rapidities has to be
symmetric for any magnetization of the system, which yields the
presence of the second energy scale in the Bethe ansatz solution of
the Kondo problem. 
This confirms the correctness of the conclusions made in my study \cite{Zv1}.

\vfill
\eject

\end{document}